\documentclass[%
aps,pra,twocolumn,superscriptaddress,showpacs,amsmath,amssymb]%
{revtex4-1}

\usepackage{hyperref}
\usepackage{graphicx}
\usepackage{bm}
\graphicspath{{./}{./figures/}}

\newcommand{\al}{{\itshape et al.} }
\usepackage{ulem,xcolor}

\newcommand{\rem}[1]{\iffalse{\color{red}\sout{#1}}\fi}
\newcommand{\add}[1]{{\iffalse\color{blue}\fi#1}}

\begin{document}
\title{Trapping of Rydberg Atoms in Tight Magnetic %
Microtraps}
\author{A.~G.~Boetes}
\affiliation{University of Amsterdam, Netherlands}
\author{R.~V.~\surname{Skannrup}} 
\email{r.u.skannrup@tue.nl}
\affiliation{Eindhoven University of Technology, Netherlands}
\author{J.~Naber} 
\affiliation{University of Amsterdam, Netherlands}
\author{S.~J.~J.~M.~F.~Kokkelmans}
\affiliation{Eindhoven University of Technology, Netherlands}
\author{R.~J.~C.~Spreeuw}
\affiliation{University of Amsterdam, Netherlands}
\email{r.u.skannrup@tue.nl}
\date{\today}

\begin{abstract}
We explore the
possibility to trap Rydberg atoms in tightly confining magnetic 
microtraps. The trapping 
frequencies for Rydberg atoms are expected to be influenced strongly by 
magnetic field gradients. We show that there are regimes where Rydberg 
atoms can be trapped. Moreover, we show that so-called magic trapping 
conditions can be found for certain states of rubidium, 
where both Rydberg atoms and ground state atoms 
have the same trapping frequencies. Magic trapping is highly beneficial 
for implementing quantum gate operations that require long operation 
times. 
\end{abstract}
\maketitle

\section{Introduction}
\label{sec:intro}
Atoms with one electron excited to a high principle quantum number 
$n$, commonly known as Rydberg atoms \cite{gallagher2005rydberg}, have 
been proposed as the basis 
for quantum simulators and quantum information processing
\cite{RevModPhys.82.2313,Muller2011}.
An idea going back to Richard Feynman 
\cite{Feynman1982}, a quantum simulator is an easily manipulated 
quantum system onto which the Hamiltonian of other quantum problems 
can be mapped. Ever since, quantum simulation and information processing 
has been driven by the promise of access to complex quantum 
systems as well as applications in quantum technology
\cite{flagship}.

In this context, Rydberg atoms attract a lot of attention due to their 
extreme properties like $n^{11}$ 
scaling of the $C_6$ Van der Waals coefficient and the blockade effect, 
providing strong interactions and the essential mechanism for 
quantum gates 
\cite{RevModPhys.82.2313,PhysRevLett.87.037901,PhysRevLett.85.2208}. 
An important issue is to achieve so-called magic trapping, identical 
trapping potentials for the ground state and the Rydberg state. Magic 
trapping conditions can suppress decoherence due to atomic 
motion during quantum 
gate protocols, much needed for high-fidelity quantum operations
\cite{WilGaeBro10,0953-4075-49-20-202001}. 
However, this is challenging to realize for alkali atoms in optical 
traps \cite{zhang2011Magic,topcu2014Divalent}.

\add{In this paper we show that magic trapping conditions can be a
chieved more easily in magnetic lattice traps \cite{WANG20161097}. Both 
ground state and Rydberg atoms can be trapped in magnetic fields arising 
from microwires on a fabricated chip 
\cite{PhysRevLett.83.3398,PhysRevLett.84.4749,Folman2002Microscopic,Reichel2002Microchip}, 
or from a patterned magnetic film giving rise to an array of microtraps 
\cite{leung2014magnetic,whitlock2009two,PhysRevA.96.013630}.}
These microtraps have very strong magnetic field gradients, 
hence are very tight, and can be arranged \add{into} different lattice 
geometries, \add{such as square or hexagonal. Field gradients can be 
particularly strong using patterned magnets. Whereas gradients above 
microwires are typically 10-100 T/m, with magnetic film chips they 
can be readily two orders of magnitude higher.}
If the blockade radius is comparable to, or larger than, 
a single trap, each trap effectively becomes a single excitation site. 
In this paper we investigate the magnetic trappability of alkali 
Rydberg atoms, and address the issue of achieving magic trapping 
conditions.

For ground 
state atoms the magnetic field can be assumed to be uniform 
across the atom. However, the large classical electron orbit radii 
of the Rydberg atoms and the large gradients of the microtraps make 
this approximation invalid.
Magnetic trapping  of Rydberg atoms in other magnetic configurations 
has been studied by other authors 
\cite{PhysRevLett.95.243001,singh2008one,PhysRevLett.95.053001,%
PhysRevA.72.053410,0953-4075-40-5-015,PhysRevLett.97.223001,%
PhysRevA.76.053417,PhysRevA.80.053410,PhysRevA.79.041403}. 
Our paper is related to the work performed by 
Mayle, Lesanovsky and Schmelcher (MLS) 
\cite{PhysRevA.80.053410,PhysRevA.79.041403}, 
\add{however, our} work is focused on the strong gradient regime 
of the microtraps, requiring a higher 
order expansion of the magnetic fields.

\add{In this work we base our calculations on $^{87}$Rb atoms, however, 
the treatment is generally applicable to other species as well.}
For a 50 kHz trap the 
oscillator length of a rubidium atom (34 nm) is much smaller than 
the rms radius of a $n=50$ electron orbit (132 nm). The strong 
magnetic field gradient then results in a magnetic field difference 
of 0.9 G across the size of the atom, resulting in an energy 
difference of about 1.3 MHz, much larger than the trapping frequency.
This paper therefore considers the effect of the spatial extent
of the electronic wave function on the trappability of the Rydberg 
atom in a magnetic trap.

\add{
In accordance with the Born-Oppenheimer approximation, we assume 
that the motion of the Rydberg electron and the atomic core can 
be separated, and that the light Rydberg electron will react instantly to any 
movement of the heavy core. We then use perturbation theory in the fine 
structure basis to find the energy of the Rydberg electron as function of the 
position of the core in the trap. These energies can be regarded as potentials 
for the core, which we call Potential Energy Surfaces (PES). We have 
expanded these potentials as harmonic traps around their respective 
minima and found feasible trapping conditions for a wide range of Rydberg states.
}

This paper is divided into six sections. In section 
\ref{sec:paramet} we provide a detailed description of the magnetic 
field configuration used in this work. In section \ref{sec:hamil} 
we provide the model Hamiltonian in Jacobi coordinates 
and discuss some of the differences to the earlier work by MLS. 
Furthermore, 
we discuss the perturbative treatment of the system. In sections 
\ref{sec:pes} and \ref{sec:trap} we discuss the outcome of the 
previous sections, with focus on trapping Rydberg atoms and magic 
trapping conditions. In section \ref{sec:conc} we conclude on our 
work.

\section{Parametrization of the magnetic traps}
\label{sec:paramet}
In the following two sections we use atomic ($\hbar=m_e=a_0=1$) units 
and summation over repeated indices
for the sake of readability. 
We model the magnetic field as a Ioffe-Pritchard configuration around 
the trap minimum
\cite{PhysRevA.74.043405}
\begin{align}
\bm{B}(\bm{x})=
\begin{pmatrix}0\\0\\\mathcal{B}\end{pmatrix}
+\mathcal{G}\begin{pmatrix}x_1\\-x_2\\0\end{pmatrix}
+\frac{1}{2}\hat{\bm{e}}_ic_{ijk}x_jx_k.
\label{eq:Bfield}
\end{align}
We shall call these terms constant $\bm{B}_c$, linear 
$\bm{B}_l$ and quadratic $\bm{B}_q$ respectively. 
The strength of the constant term is set to 3.23G. At this field the 
differential Zeeman shift between the two qubit states 
$\left|F=1,m_F=-1\right>$ and $\left|F=2,m_F=1\right>$ 
vanishes \cite{Davis2002}.

Expanding the magnetic fields to quadratic order goes beyond existing
works in literature \cite{PhysRevA.80.053410}. This is necessary 
for the systems with strong magnetic gradients we explore. 
This provides further accuracy for systems already 
investigated with linear only expansions, which can never explain axial 
trapping.

The linear term coefficient $\mathcal{G}$ provides confinement in the 
tight transverse directions. This coefficient has a value of 
900 T/m for microtraps in a hexagonal lattice \cite{leung2014magnetic}. 
In the remainder of this paper we use microtrap parameters as relevant 
for this hexagonal lattice. This is much 
greater than that of more conventional Z-wire magnetic chip 
traps with $\mathcal{G}\approx 7$ T/m \cite{naber2016magnetic}.

The curvature tensor $c_{ijk}$, which determines the strength of the 
quadratic term of the magnetic field, is symmetric under permutation 
of its indices and all partial traces vanish. This leaves 
7 independent components. For the  
microtraps the non zero components of $c_{ijk}$ are on the order of 
$10^7$ T/m\textsuperscript{2}, again much larger than for a typical 
Z-wire trap, where the non zero components are on the order of 10-100 
T/m\textsuperscript{2}.

Choosing the Coulomb gauge we find the vector 
potential corresponding to Eq. \eqref{eq:Bfield}
\begin{align}
\bm{A}(\bm{x})=\frac{\mathcal{B}}{2}
\begin{pmatrix}
-x_2\\x_1\\0
\end{pmatrix}
+\mathcal{G}\begin{pmatrix}
0\\0\\x_1x_2
\end{pmatrix}
+\frac{1}{8}\hat{\bm{e}}_i\epsilon_{ijk}c_{jlm}x_lx_mx_k,
\end{align}
with $\epsilon_{ijk}$ the fully antisymmetric Levi-Civita tensor.
We retain the naming convention from the magnetic field, i.e. the 
curl of the 'linear' term of the vector potential corresponds to the 
linear term of the magnetic field 
\mbox{$\nabla\times\bm{A}_l=\bm{B}_l$} etc.
It is convenient to define "residual terms" for the 
magnetic field and the vector potential respectively, as follows
\begin{align}
\tilde{\bm{B}}(\bm{R},\bm{r})=&\bm{B}(\bm{R}+\bm{r})-\bm{B}(\bm{R})
-\bm{B}(\bm{r})\\
\tilde{\bm{A}}(\bm{R},\bm{r})=&\bm{A}(\bm{R}+\bm{r})-\bm{A}(\bm{R})
-\bm{A}(\bm{r}).
\end{align}
Note that these do not describe the fields at any position, but merely 
express the difference between the sum of fields at two positions 
and the field at the sum of those two positions. 

\section{Hamiltonian and perturbation terms}
\label{sec:hamil}
\begin{figure}[h]
\centering
\rotatebox{-90}{
\includegraphics[width=.6\columnwidth]{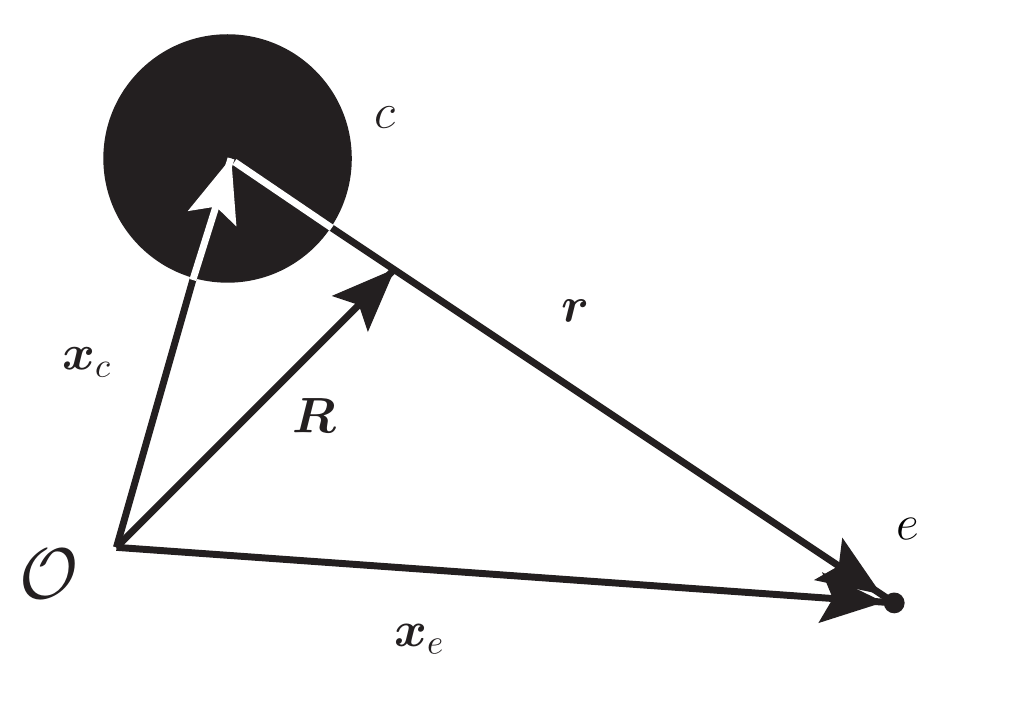}}
\caption{\itshape Schematic of the coordinates used in this paper. 
$\mathcal{O}$ denotes the origin. $\bm{x}_e$ and $\bm{x}_c$ are the 
position vectors of the electron and core respectively in the 
Ioffe-Pritchard frame. $\bm{r}$ and $\bm{R}$ denotes the relative and 
center of mass coordinates respectively. Since the mass is almost 
entirely contained in the core we approximate $\bm{R}$ with 
$\bm{x}_c$.
\label{fig:jaccoo}}
\end{figure}
Our approach builds on a spin and minimal coupling scheme for the 
valence electron with position $\bm{x}_e$, momentum $\bm{p}_e$ and 
mass $m_e=1$ and the core with position $\bm{x}_c$, momentum $\bm{p}_c$ 
and mass $M$ in an external magnetic field. We 
reexpress this Hamiltonian using Jacobi coordinates 
(\mbox{$(x,y,z)^T=\bm{r}=\bm{x}_e -\bm{x}_c$}, 
\mbox{$(X,Y,Z)^T=\bm{R}=\bm{x}_e/M+\bm{x}_c$}, see Fig. 
\ref{fig:jaccoo}). 
Since the mass ratio is large, about $1.6\times10^5$ for 
\textsuperscript{87}Rb, 
we identify the core $\bm{x}_c$ and center of mass coordinate $\bm R$ as 
an approximation. This leaves us with the Hamiltonian
\begin{align}
\mathcal{H}=&\mathcal{H}_\mathrm{ff}
+\bm{A}\left(\bm{R}+\bm{r}\right)\cdot\bm{p}
+\bm{S}\cdot \bm{B}\left(\bm{R}+\bm{r}\right)\nonumber\\
&+\frac{\bm{P}^2}{2M}
+\frac{1}{M}\left[\bm{A}\left(\bm{R}+\bm{r}\right)
-\bm{A}\left(\bm{R}\right)\right]\cdot\bm{P}\\
&+\frac{1}{2}\bm{A}^2\left(\bm{R}+\bm{r}\right)
+\frac{g_I}{2}\bm{I}\cdot\bm{B}\left(\bm{R}\right),\nonumber
\end{align}
with $\mathcal{H}_\mathrm{ff}$ being the (field free) fine structure 
Hamiltonian, $\bm{P}$ the center of mass 
momentum operator, $\bm{S}$ the electron 
spin, $\bm{I}$ the nuclear spin and $g_I$ the nuclear Land\'e 
g-factor.
We apply degenerate perturbation theory in the 
fine structure basis $\left|\kappa\right>=\left|nLjm_j\right>$
to this Hamiltonian, coupling to all states within one $L$-manifold.

At this point, in previous work 
\cite{PhysRevA.80.053410,PhysRevA.79.041403}
a unitary transformation is  applied:
$U_\mathrm{lit}=\exp(-i(\bm{B}_c\times \bm{r})\cdot \bm{R}/2)
=\exp(i\bm{A}_c(\bm{R})\cdot \bm{r})$. 
This removes, what they consider the most dominant perturbation terms 
in the linear field approximation, 
$\bm{A}_c^2(\bm{R})/2$ and $-\bm{A}_c(\bm{R})\cdot \bm{p}$, which 
are of similar magnitude but opposite sign, making the perturbative 
treatment more robust. However, this transformation complicates 
the diamagnetic terms unnecessarily. 
If we consider Eq. (20) in  Ref. \cite{PhysRevA.80.053410} along the 
line $X=Y$ we obtain 
$E^{(2)}_\kappa(\bm{R})\approx C_z\mathcal{G}^2X^4$ with 
$C_z\approx -1/2$, within 10\% for $35\leq n\leq 45$ and $l\leq 4$. 
But if we consider the term 
$\bm{A}^2_l(\bm{R})/2=\mathcal{G}^2X^2Y^2/2$, which is implicitly 
neglected by MLS, we find the same value but with opposite 
sign along $X=Y$. Thus the main Rydberg contribution is countered 
by a neglected term. 

Instead we use a more general unitary 
transformation that does not rely on any explicit field to be 
introduced, and is inspired by the previous 
\begin{align}
U=\exp\left(i\bm{A}(\bm{R})\cdot \bm{r}\right). \label{eq:tran}
\end{align}
We apply this to the Hamiltonian: $H=U\mathcal{H}U^\dagger$.
The transformation \eqref{eq:tran} removes the 
terms $\bm{A}^2(\bm{R})/2$ and $-\bm{A}(\bm{R})\cdot \bm{p}$, 
in their entirety in contrast to the standard transformation 
$U_\mathrm{lit}$.

The resulting Hamiltonian can be split into four parts
($H=H_{R,r}+H_R+H_r+H_{r,P}$)
according to 
their dependence on the Jacobi operators
\begin{align}
H_{R,r}=&
H_\mathrm{ff}+\left(\bm{S}+\frac{1}{2}\bm{L}_r\right)\cdot \bm{B}(\bm{R})
+\frac{1}{2}\tilde{\bm{A}}^2(\bm{R},\bm{r})\nonumber\\
+&\bm{A}(\bm{r})\cdot\tilde{\bm{A}}(\bm{R},\bm{r})
+H_\mathrm{small},
\label{eq:HRr}
\\
H_{R}=&
\frac{\bm{P}^2}{2M}+\frac{1}{2}g_I \bm{I}\cdot \bm{B}(\bm{R}),
\\
H_{r}=&\nonumber
\left(\bm{B}_l(\bm{r})+\bm{B}_q(\bm{r})\right)\cdot \bm{S}
+\left(\bm{A}_l(\bm{r})+\bm{A}_q(\bm{r})\right)\cdot \bm{p}\\
+&\frac{1}{2}\bm{A}^2(\bm{r}),
\\
H_{r,P}=&
\frac{1}{M}\left[\bm{A}(\bm{R}+\bm{r})-\bm{A}(\bm{R})
+\nabla_R\left(\bm{A}(\bm{R})\cdot\bm{r}\right)\right]\cdot\bm{P},
\end{align}
with $H_\mathrm{small}$ 
\footnote{
We have neglected a number of terms in Eq. \eqref{eq:HRr}, part of the 
perturbation Hamiltonian
\begin{align*}
H&_\mathrm{small}=
\bm{S}\cdot\tilde{\bm{B}}_q(\bm{R},\bm{r})
+\frac{\alpha^2}{2r}\frac{dV_l(r)}{dr}(\bm{A}(\bm{R})\times\bm{r})
\cdot \bm{S}\\
+&\frac{i}{2}\mathcal{G}\big(X\left[H_\mathrm{ff},yz\right]
+Y\left[H_\mathrm{ff},xz\right]\big)\\
+&\frac{i}{2}\partial^R_j\bm{A}_{q,k}
(\bm{R})\left[H_\mathrm{ff},r_jr_k\right]
+\bm{A}^{(2)}_q(\bm{R},\bm{r})\cdot\bm{p},\label{eq:negl}
\end{align*}
which have all been estimated to give only minor contributions.
With the exception of $\bm{S}\cdot\tilde{\bm{B}}_q(\bm{R},\bm{r})$ 
term, all have higher order than 3 in the relative coordinates.
}
collecting some terms we can neglect in perturbation theory, and 
$\bm{L}_{r}$ being the electron angular momentum operator.
For the ground state, only the $\bm{B}(\bm{R})$ term of Eq. 
\eqref{eq:HRr} contributes significantly to the energy, as 
$\langle|\bm{r}|\rangle\approx 0$, and the magnetic trapping field 
is assumed to 
be constant across the atom. This is sharply contrasted for large
Rydberg states, where terms dependent on $\bm{r}$ become important, 
since $\langle|\bm{r}|\rangle\propto n^2$ is large, and the terms
\begin{align}
\frac{1}{2}\tilde{\bm{A}}^2(\bm{R},\bm{r})+\bm{A}(\bm{r})\cdot
\tilde{\bm{A}}(\bm{R},\bm{r})+H_r,
\label{eq:Hryd}
\end{align}
which we call the "Rydberg term",
become important. These terms \add{are mostly extra terms compared to 
the MLS approach, and together} constitute the Rydberg specific part of 
the Hamiltonian.

We work in a frozen gas setting where $P/M\approx 0$. This has the 
direct consequence that we can neglect the $H_{r,P}$. Further more,
this setting is well explored with the Born-Oppenheimer 
approximation, where the electrons are assumed to react instantly to 
any core movement.
In accordance with the Born-Oppenheimer approximation we assume the
eigenstates to be product states of a $\bm{r}$ 
dependent part and a $\bm{R}$ dependent part
\begin{align}
|\psi\rangle=\left|\psi_{\bm{r}}\right>\left|\psi_{\bm{R}}\right>
=\sum_\kappa c_\kappa\left|\kappa\right>\left|\psi_{\bm{R}}\right>,
\end{align}
where $c_{\kappa}$ are expansion coefficients for $\psi_{\bm{r}}$ 
in the fine structure basis.

By applying the electronic parts (i.e. the parts 
dependent on the relative coordinate) of the Hamiltonian, we find 
an energy dependent on the core position $\bm{R}$
\begin{align}
(H_{R,r}+H_r)\left|\psi\right>
=E(\bm{R})\left|\psi\right>.
\label{eq:HonPsi}
\end{align}
We specifically use degenerate perturbation theory to find 
the electronic energies $E(\bm{R})$ at any given core position 
$\bm{R}$.
We use a set of all fine structure states 
$|\kappa\rangle=|nLjm_j\rangle$ (the eigenstates of $H_\mathrm{ff}$),
within one $n$, $L$- manifold, as 
basis for our perturbative treatment, as the energy 
contribution from the fine structure Hamiltonian is, by far, most 
dominant.
We have found that coupling between states with different $n$ or $L$ 
quantum numbers is not significant for the parameter space we are 
considering, and we have not included this in our model.

The complexity of this computation can be greatly reduced by 
carefully examining and understanding the couplings between different 
states. The expressions become quite simple and $S_{1/2}$ states can 
be solved analytically.
We include 
mixing between different $j$ states within one $L$ manifold, as they 
are sufficiently close in energy for the principal quantum numbers of 
interest.

Since the energy in Eq. \eqref{eq:HonPsi} is dependent on the core 
position $\bm{R}$ we interpret it as a potential and construct a 
total potential $W$ seen by the core
\begin{align}
\nonumber
\langle\psi_{\bm{r}}|H|\psi\rangle=&\left[H_R+E(\bm{R})\right]|\psi_{\bm{R}}\rangle\\
=&\left[T_R+W(\bm{R})\right]|\psi_{\bm{R}}\rangle.
\end{align}
We call these potentials $W(\bm{R})$ Potential Energy Surfaces (PES).
\add{Since the mictrotraps are designed to trap ground state atoms with 
only little spatial extent, it can be expected that trapping is mostly 
provided by the unperturbed Hamiltonian. However, there are exceptions 
leading to anti-trapping, as we will explain below.}

\section{Potential energy surfaces}
\label{sec:pes}
\begin{figure}[ht]
\centering
\includegraphics[width=\columnwidth]{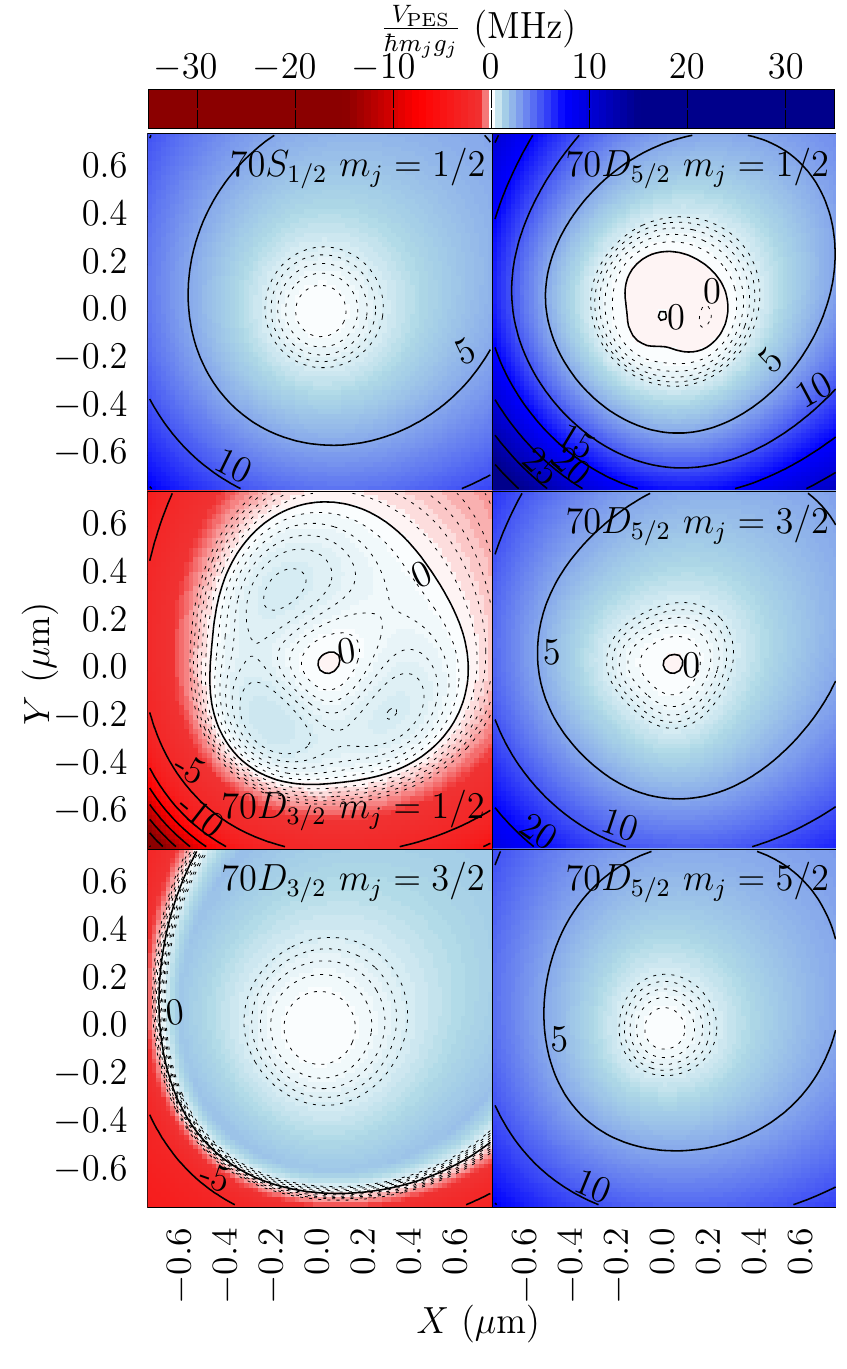}
\caption{\itshape Potential Energy Surfaces (PES) for the different 
angular states with $n=70$ in the Ioffe-Pritchard plane ($Z=0$) scaled 
by $m_jg_j$. 
These represent results for a typical magnetic potential of the 
hexagonal magnetic lattice.
Solid contours at every 5 MHz and dashed contours at every 0.2 MHz 
in the interval -1 to 1 (limits included) are shown. 
Near the origin we observe positive curvature, meaning that the state 
is trappable, for all but the $70D_{5/2}$ with $m_j=1/2$ state; where 
a small bump indicates a crossover to a Mexican Hat type potential. 
At higher $n$ this will be more pronounced.
For both $70D_{3/2}$ states we observe strong downwards gradients for 
large $|R|$, due to the strong influence of the diamagnetic term. 
The noticeable asymmetry in the plots is due to the fact that the 
coordinates are rotated with respect to the atom chip surface. 
The $X=Y$ direction is normal to the surface.
\label{fig:PES}}
\end{figure}
\begin{figure}[ht]
\centering
\includegraphics[width=\columnwidth]{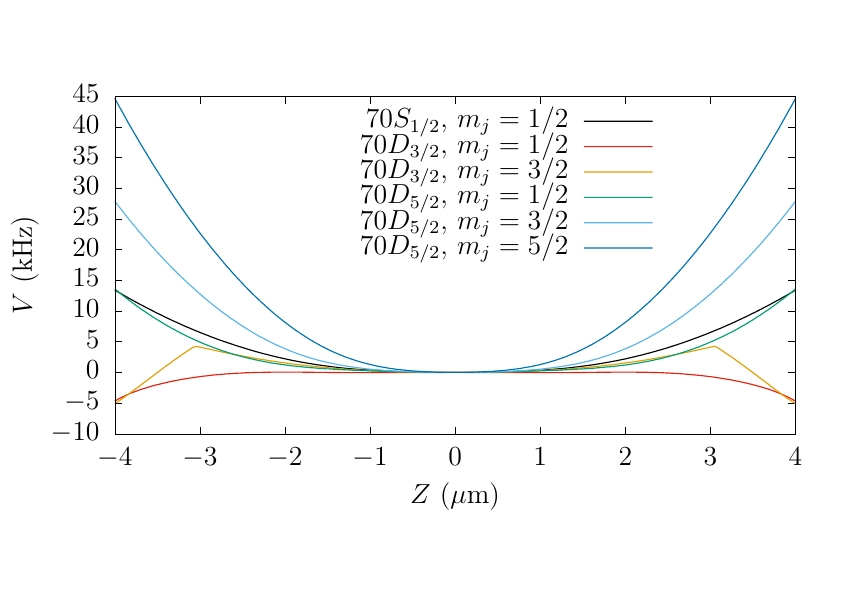}
\caption{\itshape
Trapping potential in the $Z$ direction at $X=Y=0$ for the different 
angular states with $n=70$. Parameters of the hexagonal lattice 
microtraps have been used \cite{leung2014magnetic}. 
Trapping along this direction is much weaker than in the $X$, $Y$ 
plane. All states shown have local, albeit weak, minima near $Z=0$.
\label{fig:Zpot}}
\end{figure}

We have calculated PES states that are reachable via a standard 
two-photon excitation process ($S$- and $D$ states) from the rubidium 
ground state. 
We consider only states where $n\leq 80$ in order to 
keep perturbations small compared to the fine structure energies and 
to not break the Born-Oppenheimer approximation. For $nD_j$ states 
mixing becomes significant when $n>80$ and the finestructure states 
are no longer good quantum states. Thus the results for 
$nD_j$ with $n>80$ are unreliable. 
In the remainder of this paper we no longer use atomic units, but 
rather SI units.

\begin{figure*}[ht]
\centering
\includegraphics[width=\columnwidth]{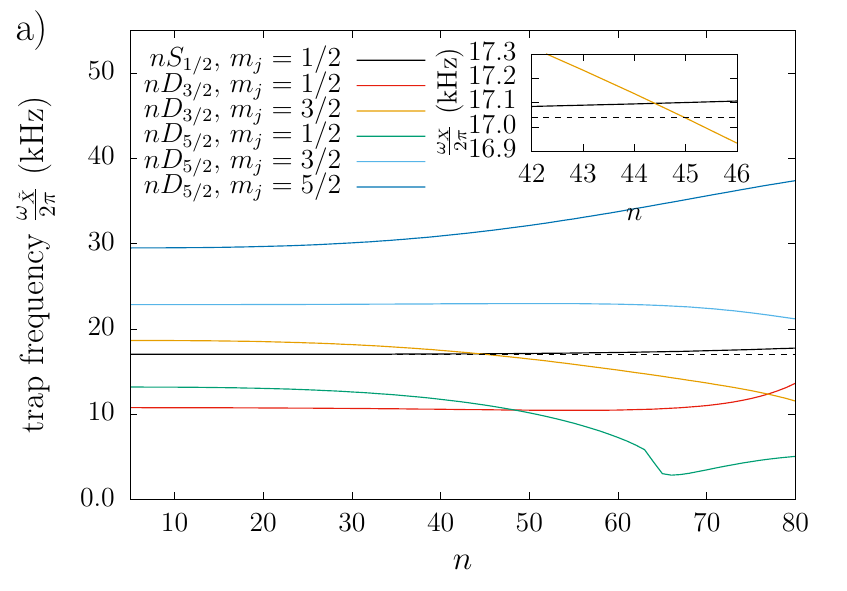}
\includegraphics[width=\columnwidth]{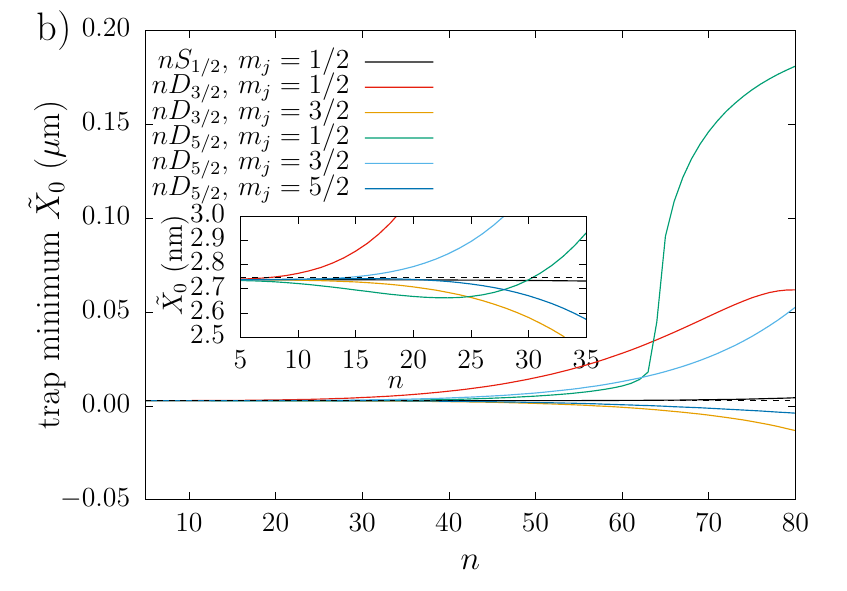}
\caption{\itshape 
(a) Radial trap frequency and
(b) position of the local potential minimum as a function of $n$, both 
in the $\tilde{X}$ direction. 
The shift of the trap minimum away from the origin occurs gradually 
for low $n$ but becomes very rapid for larger $n$. The rapid shift 
indicates crossover to a Mexican hat type potential.
Dashed lines indicates the value of the 
$5S_{1/2}$, $m_j=1/2$ state in all cases.
Similar results are found in the 
$\tilde{Y}$ direction; whereas in the $Z$ direction the trap is 
an order of magnitude weaker.
Of special interest is the $45D_{3/2}$, $m_j=3/2$ state, where 
the trapping frequency is very close to that of the ground 
state. This and the 
small value of $X_0$ lead to magic trapping conditions for the 
$45D_{3/2}$, $m_j=3/2$ state. 
\label{fig:eff_trap}}
\end{figure*}

In Fig. \ref{fig:PES} we present the PES 
for the $70L_j$ states with all different positive $m_j$ up to a 
distance of 0.75 $\mu$m from the trap center in the radial plane 
($Z=0$). These have been rescaled by $m_jg_j$ to make them comparable.
The approximate symmetry with respect to the $X=Y$ line is due to 
this line being normal to the chip surface.
In Fig. \ref{fig:Zpot} we see the $Z$ dependence of the PES for the same 
states to a distance of 4 $\mu$m from the trap center along the $X=Y=0$ 
line.
The choice of the $70L_j$ states is motivated by being well within the 
limits of our methods while the high $n$ makes the Rydberg specific 
contributions clearly visible.
This is seen in the strong dependence on the angular 
state, which is not evident for $n<40$. 
When going to even higher $n$ these effects become more pronounced and  
we eventually loose trapping for the $nD_{3/2}$ potentials, whereas 
the $nD_{5/2}$ states transition to quartic trapping potentials. 

The $nS_{1/2}$ state trapping potentials do remain fairly 
similar to that of the ground state, not surprising as the electron 
is more tightly confined near the core.

The $70D_{3/2}$ states stand out among the PES by being antitrapping on 
the micrometer scale in both the $Z=0$ plane and along the $X=Y=0$ line.
The PES of these states are more strongly influenced by the diamagnetic 
terms in the Hamiltonian, leading to both the antitrapping behavior 
and the structure in the positive potential region of 
the $m_j=1/2$ state, by coupling to $(j$, $m_j)$-states of different 
angular symmetry. This structure makes the state unsuitable 
for quantum simulation but shows the importance of the Rydberg 
nature of the atom to the PES.

We see a small bump near the origin in the PES of the 
$70D_{5/2}$, $m_j=1/2$ state. For higher $n$ this bump becomes 
a regular peak turning the potential a Mexican hat shape.

In our analysis of the PES of the $n=35$, $45$, $55$, $65$, and $75$ 
states, we fitted a polynomial to the contributions of
the Rydberg specific terms in Eq. \eqref{eq:Hryd}.
This showed that, though highly state dependent, the effect of the 
Rydberg terms can be reduced to an offset and an $R^2$ dependent term 
for the $nS_{1/2}$ and $nD_{5/2}$ states. When $n>73$, however, this 
simple picture fails for the $nD_{3/2}$ states, and higher order terms 
are needed to describe the behavior. 

We predict that one can encounter this effect in spectroscopic 
measurements, even for weaker traps, as long as 
the magnetic field is well described by the second order expansion 
of our model.

Our results show that in general, the PES of the $nS_{1/2}$ and 
$nD_{5/2}$ states are always trapping on micrometer length scales for 
atoms with $n<80$. 
The $nD_{3/2}$ states also show trapping PES, but for $n>50$ 
the PES become of antitrapping nature on the micrometer scale, and we only 
observe local trapping on sub micrometer scales,
and rich structure appearing near the center of the PES, 
when $m_j=1/2$ and $n>60$. 

\section{Trapping conditions}
\label{sec:trap}
We now analyze the trapping conditions for different 
Rydberg states as function of principal quantum number $n$ and 
angular state. We investigate in particular whether Rydberg states 
with trapping conditions identical to the ground state conditions 
can be found. 
This is particularly relevant for the implementation of 
various quantum information protocols based on Rydberg interactions. 
Rydberg atoms and ground state atoms experience different 
trapping potentials, which leads to motional decoherence.
We denote a Rydberg atom in internal state 
$\left|r\right>=\left|nLjm_j\right>$ in a certain motional state 
$\left|\nu\right>$ by $\left|r,\nu\right>$.
During the Rydberg 
excitation that same motional state $\left|\nu\right>$ will be a 
non-stationary state in the Rydberg trap. When de-exciting the atom the 
motional state will have changed under time evolution and no longer be 
identical to $\left|\nu\right>$.
Therefore it is of great interest if we can suppress 
this decoherence  mechanism by realizing conditions of magic trapping, 
where ground and Rydberg state atoms would experience identical 
trapping potentials.

First, we define the rotated coordinates away from the surface of 
the chip
$\tilde{X}=\sqrt{\frac{1}{2}}(X+Y)$ 
and $\tilde{Y}=\sqrt{\frac{1}{2}}(X-Y)$ parallel to the surface. 
\begin{align}
V(\tilde{X},\tilde{Y},Z)
\nonumber
=&C+\frac{1}{2}m\omega^2_{\tilde{X}} (\tilde{X}-\tilde{X}_0)^2
+\frac{1}{2}m\omega^2_{\tilde{Y}} (\tilde{Y}-\tilde{Y}_0)^2\\
&+\frac{1}{2}m\omega_Z^2 (Z-Z_0)^2,
\label{eq:zero_exp}
\end{align}
where $C$ is some constant offset, $(\tilde{X}_0,\tilde{Y}_0,Z_0)$ 
is the minimum 
position of the trap and $\omega_i$ is the local trap frequency around 
the minimum in the $\tilde{R}_i$ direction.
This is a good approximation over a distance of $0.3 \mu m$ from the 
trap center, many times larger than the trap oscillator length. 
Analysis of the traps show that all states have tens or hundreds of 
trap levels. With energy much lower than the potential walls, the center 
of mass will behave as in an infinitely deep potential. The lowest 
number of trap levels are found for the $nD_{3/2}$, $m_j=1/2$ states, 
consistent with the antitrapping long range behavior.

Our analysis shows that the trap behaves harmonically for $n\leq 50$ 
and does not deviate significantly from the $n=5$ trapping potential 
for any given angular state, see Fig. \ref{fig:eff_trap}.
In particular the changes in $nS_{1/2}$ state  trapping 
frequency and minimum position remain insignificant over 
the whole $n$-range considered.

\begin{figure}[ht]
\centering
\includegraphics[width=\columnwidth]{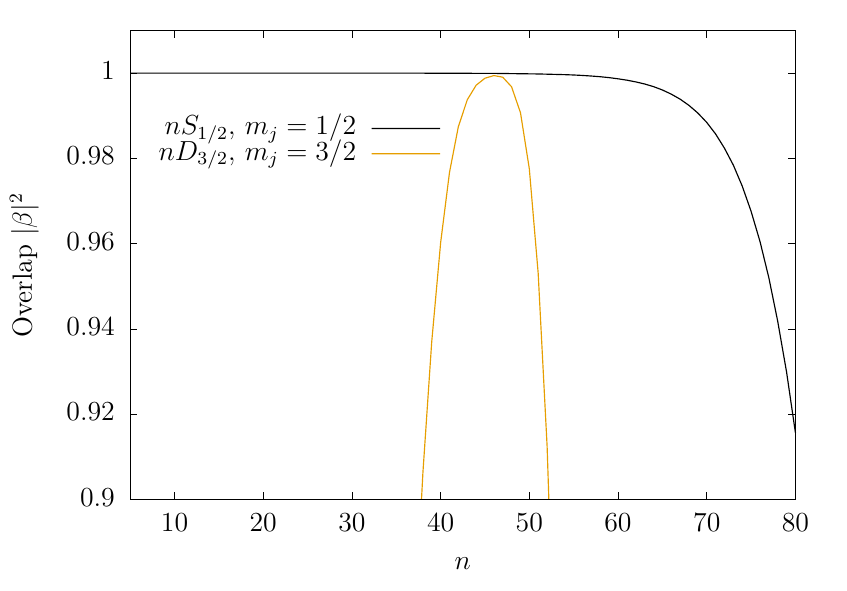}
\caption{\itshape 
Overlap of the center of mass ground state after 10 $\mu$s spent
in an excited state, taking all directions into account. 
We bring the system into the Rydberg state and 
de-excite after 10 $\mu$s, we ignore all other decoherence effects like 
spontaneous emission.
This determines the probability of finding 
the center of mass particle in the ground state after spending 
some time in a given excited electronic state. 
\label{fig:pur}
}
\end{figure}

The $nD_{3/2}$ states show strong dependence on $n$ and $m_j$. In the 
$m_j=1/2$ states we find the trap bottom shift away from the origin 
but the effective trap frequency remains relatively stable until 
$n=70$. 
The $m_j=3/2$ states, show consistently decreasing trap frequencies 
but the trap minima remain fairly centered.
As seen from Fig. \ref{fig:eff_trap} 
magic trapping conditions, i.e. effective trapping similar to that of 
the ground state, are present for the $nD_{3/2}$, $m_j=3/2$ states 
around $n=45$. The effective trap frequency 
of these states are equal to that of the ground state in the $\tilde{X}$
direction.
This means that the Rydberg excitation cycle can be performed with 
minimal motional decoherence. Since the PES for these states
are nearly identical to that for the ground state, the CM wave function 
remains unchanged after the Rydberg excitation, see Fig 
\ref{fig:pur}. 

The $nD_{5/2}$ states show rich behavior in the parameters of the 
effective trapping.
The trap frequencies of states with $m_j=1/2$ drop to around a 
quarter of the $n=5$ value at $n=66$. The minimum of the traps shift 
position away from the origin, very rapidly for $63\leq n\leq 65$.
Inspection of the PES shows that this is a crossover to a Mexican Hat 
type potential in the $Z=0$ plane.
The $m_j=3/2$ state trapping frequencies remain stable, at around twice 
the value of the ground state trap frequency, across the entire 
$n$-interval considered, with a slight decrease for very large 
$n$. The minimum position shifts away from the origin.
The $m_j=5/2$ states show large, increasing trapping frequency, but no 
significant change in trap minimum position.
The $nD_{5/2}$ states are not suitable for procedures requiring 
trap frequencies comparable to those of the ground state.

The trapping conditions that were found for the $\tilde{X}$ direction in
Fig. \ref{fig:eff_trap} for the $45D_{3/2}$, $m_j=3/2$ and nearby states 
(and similar conditions in the $\tilde{Y}$ and $Z$ directions) are 
expected to strongly suppress motional decoherence in any gate protocol 
involving Rydberg excitation and de-excitation. A full analysis of gate 
fidelities should take into account the photon recoils upon
(de-)excitation, considering also that these tight magnetic traps are 
in the Lamb-Dicke limit\cite{Wineland98}. For the highest 
fidelities the anharmonicity of the traps may also play a role. Such 
full analysis of fidelities is beyond the scope of this paper.

As a first indicator, we have projected the motional ground state of 
the $5S_{1/2}$, $m_j=1/2$ electronic state, denoted by 
$\left|g,0\right>$,
onto the motional states of an electronically excited state,
\begin{align}
\left|\psi\right>
=&\sum_\nu\left|r,\nu\right>
\left<g,0\middle|r,\nu\right>
=\sum_\nu c_\nu\left|r,\nu\right>,
\end{align}
with $\nu=\{\nu_{\tilde{X}},\nu_{\tilde{Y}},\nu_{Z}\}$ being the 
motional quantum numbers in the indexed directions.
We time evolve this projection 
$\left|\psi(t)\right>=\exp(i H t/\hbar)\left|\psi\right>$ and 
calculate the time delayed overlap
\begin{align}
\beta_t=&\left<\psi(0)\middle| \psi(t)\right>\nonumber\\
=&\sum_\nu |c_\nu|^2
\exp\left[-i
\left(\nu+\frac{3}{2}\right)\omega t\right],
\end{align}
where the coefficients $c_\nu$ are calculated using second order 
perturbation theory,and define the overlap of the state as 
$|\beta|^2=|\beta_{10\mu s}|^2$. 
Thus $1-|\beta|^2$ 
represents the probability of finding the atom in a different motional 
state after a 10 $\mu$s evolution time. 

In Fig. \ref{fig:pur} we see that this overlap reaches 0.9994, for the 
$46D_{3/2}$, $m_j=3/2$ state, comparable to that of the 
$nS_{1/2}$, $m_j=1/2$ states, which have overlaps exceeding 0.999 for 
$n\leq 58$.
We have identified two angular configurations that are comparably good, 
around $n=46$. Experiments therefore can choose between two angular 
states and therefore also between symmetric and directional blockade 
regions. The reason that the optimal overlap is found for 
$46D_{3/2}$, whereas Fig \ref{fig:eff_trap} would suggest $n=45$ is due 
to a small shift of the trap minimum position.

This overlap allows for minimizing decoherence
due to changes in effective trapping, making sure the 
effect of the excited state trap is not the limiting factor.
This should be compared to losses due to other sources, which will be 
dominant, in particular spontaneous emission and transitions driven by
black body radiation, with a life time of about 50 $\mu$s for the 
$n=45$ states \cite{PhysRevA.79.052504}.

\section{Conclusion and outlook}
\label{sec:conc}
We have studied Rydberg atoms in magnetic microtraps 
described by a second order expansion of the magnetic field. 
The magnetic microtraps are much tighter and have much stronger 
field gradients than more commonly used traps, such as Z-wire traps.
This enhances the effects of the trap on the spatially extended Rydberg 
atoms. 

Our work confirms the findings by Mayle \al \cite{PhysRevA.80.053410}, 
that Rydberg atoms can indeed be magnetically trapped,
\add{ and we have extended their model by including several new 
terms in the Hamiltonian,} most 
importantly thediamagnetic term mixing the relative and center of 
mass coordinates. 
These terms constituted an unknown contribution to the trapping 
potentials of Rydberg atoms that, while negligible in weaker traps, 
are important in the context of microtraps.
We have, however, also found the 'Rydberg term' of Ref. 
\cite{PhysRevA.80.053410} to be almost 
zeroed by some of the \add{additional} terms.

We found that trapping of Rydberg atoms is possible for both 
$S$-states and $D$-states, but for high $n$ 
effective trapping potentials become distorted,
due to the anisotropic nature of the Rydberg contributions 
and the increased contribution from the diamagnetic terms.

We have found near-magic trapping conditions with more than 
99\% overlap for $nS$ states with $n<70$ and $nD_{3/2}$ 
states with $m_j=3/2$ and $43\leq n\leq 49$, with the highest overlap 
for the $n=46$ state.
This provides a choice between the two angular states, and therewith 
the angular dependence of the interaction.
With magic trapping states available, a Rydberg equivalent of the 
M{\o}lmer-S{\o}rensen gate 
\cite{PhysRevLett.82.1971,PhysRevA.62.022311,PhysRevLett.86.3907}, 
relying on such conditions, could be possible. Such a gate 
implementation will be of great value for quantum simulation 
and processing, and demands further research.

We have found that the spatially extended nature of Rydberg atoms has 
significant effects in the microtraps, and results significant 
modifications of the trapping potentials of the center of mass. 
In particular we have observed a strong $n$-dependence 
of the center of mass trapping potentials, with 
shallow trapping for $nD_{3/2}$ states and quartic trapping of 
$nD_{5/2}$ states.

Further research should consider the effect of the trap on the 
electronic wave function and, in turn, the effect on the Rydberg-Rydberg 
interaction. 

Finally we remark that the methods employed in this work 
can readily be adapted to model other isotopes or elements. 
By adjusting the magnetic field parameters, we can model other 
magnetic trap configurations.

\begin{acknowledgments}
We would like to thank Ben van den Linden van den Heuvel 
and Ren\'e Gerritsma 
for feedback and discussions.
This research was financially supported by the Foundation for 
Fundamental Research on Matter (FOM), and by the Netherlands 
Organization for Scientific Research (NWO). We also acknowledge 
the European Union H2020 FET Proactive project RySQ (grant N. 
640378).
\end{acknowledgments}


\begin{thebibliography}{37}%
\makeatletter
\providecommand \@ifxundefined [1]{%
 \@ifx{#1\undefined}
}%
\providecommand \@ifnum [1]{%
 \ifnum #1\expandafter \@firstoftwo
 \else \expandafter \@secondoftwo
 \fi
}%
\providecommand \@ifx [1]{%
 \ifx #1\expandafter \@firstoftwo
 \else \expandafter \@secondoftwo
 \fi
}%
\providecommand \natexlab [1]{#1}%
\providecommand \enquote  [1]{``#1''}%
\providecommand \bibnamefont  [1]{#1}%
\providecommand \bibfnamefont [1]{#1}%
\providecommand \citenamefont [1]{#1}%
\providecommand \href@noop [0]{\@secondoftwo}%
\providecommand \href [0]{\begingroup \@sanitize@url \@href}%
\providecommand \@href[1]{\@@startlink{#1}\@@href}%
\providecommand \@@href[1]{\endgroup#1\@@endlink}%
\providecommand \@sanitize@url [0]{\catcode `\\12\catcode `\$12\catcode
  `\&12\catcode `\#12\catcode `\^12\catcode `\_12\catcode `\%12\relax}%
\providecommand \@@startlink[1]{}%
\providecommand \@@endlink[0]{}%
\providecommand \url  [0]{\begingroup\@sanitize@url \@url }%
\providecommand \@url [1]{\endgroup\@href {#1}{\urlprefix }}%
\providecommand \urlprefix  [0]{URL }%
\providecommand \Eprint [0]{\href }%
\providecommand \doibase [0]{http://dx.doi.org/}%
\providecommand \selectlanguage [0]{\@gobble}%
\providecommand \bibinfo  [0]{\@secondoftwo}%
\providecommand \bibfield  [0]{\@secondoftwo}%
\providecommand \translation [1]{[#1]}%
\providecommand \BibitemOpen [0]{}%
\providecommand \bibitemStop [0]{}%
\providecommand \bibitemNoStop [0]{.\EOS\space}%
\providecommand \EOS [0]{\spacefactor3000\relax}%
\providecommand \BibitemShut  [1]{\csname bibitem#1\endcsname}%
\let\auto@bib@innerbib\@empty
\bibitem [{\citenamefont {Gallagher}(2005)}]{gallagher2005rydberg}%
  \BibitemOpen
  \bibfield  {author} {\bibinfo {author} {\bibfnamefont {T.}~\bibnamefont
  {Gallagher}},\ }\href@noop {} {\emph {\bibinfo {title} {Rydberg Atoms}}},\
  Cambridge Monographs on Atomic, Molecular and Chemical Physics\ (\bibinfo
  {publisher} {Cambridge University Press},\ \bibinfo {year}
  {2005})\BibitemShut {NoStop}%
\bibitem [{\citenamefont {Saffman}\ \emph {et~al.}(2010)\citenamefont
  {Saffman}, \citenamefont {Walker},\ and\ \citenamefont
  {M\o{}lmer}}]{RevModPhys.82.2313}%
  \BibitemOpen
  \bibfield  {author} {\bibinfo {author} {\bibfnamefont {M.}~\bibnamefont
  {Saffman}}, \bibinfo {author} {\bibfnamefont {T.~G.}\ \bibnamefont {Walker}},
  \ and\ \bibinfo {author} {\bibfnamefont {K.}~\bibnamefont {M\o{}lmer}},\
  }\href {\doibase 10.1103/RevModPhys.82.2313} {\bibfield  {journal} {\bibinfo
  {journal} {Rev. Mod. Phys.}\ }\textbf {\bibinfo {volume} {82}},\ \bibinfo
  {pages} {2313} (\bibinfo {year} {2010})}\BibitemShut {NoStop}%
\bibitem [{\citenamefont {M{\"{u}}ller}\ \emph {et~al.}(2011)\citenamefont
  {M{\"{u}}ller}, \citenamefont {Haakh}, \citenamefont {Calarco}, \citenamefont
  {Koch},\ and\ \citenamefont {Henkel}}]{Muller2011}%
  \BibitemOpen
  \bibfield  {author} {\bibinfo {author} {\bibfnamefont {M.~M.}\ \bibnamefont
  {M{\"{u}}ller}}, \bibinfo {author} {\bibfnamefont {H.~R.}\ \bibnamefont
  {Haakh}}, \bibinfo {author} {\bibfnamefont {T.}~\bibnamefont {Calarco}},
  \bibinfo {author} {\bibfnamefont {C.~P.}\ \bibnamefont {Koch}}, \ and\
  \bibinfo {author} {\bibfnamefont {C.}~\bibnamefont {Henkel}},\ }\href
  {\doibase 10.1007/s11128-011-0296-0} {\bibfield  {journal} {\bibinfo
  {journal} {Quantum Information Processing}\ }\textbf {\bibinfo {volume}
  {10}},\ \bibinfo {pages} {771} (\bibinfo {year} {2011})}\BibitemShut
  {NoStop}%
\bibitem [{\citenamefont {Feynman}(1982)}]{Feynman1982}%
  \BibitemOpen
  \bibfield  {author} {\bibinfo {author} {\bibfnamefont {R.~P.}\ \bibnamefont
  {Feynman}},\ }\href {\doibase 10.1007/BF02650179} {\bibfield  {journal}
  {\bibinfo  {journal} {International Journal of Theoretical Physics}\ }\textbf
  {\bibinfo {volume} {21}},\ \bibinfo {pages} {467} (\bibinfo {year}
  {1982})}\BibitemShut {NoStop}%
\bibitem [{fla(2017)}]{flagship}%
  \BibitemOpen
  \href {https://ec.europa.eu/newsroom/document.cfm?doc_id=42721} {\enquote
  {\bibinfo {title} {Quantum technologies flagship intermediate report},}\ }
  (\bibinfo {year} {2017}),\ \bibinfo {note}
  {https://ec.europa.eu/digital-single-market/en/news/intermediate-report-quantum-flagship-high-level-expert-group}\BibitemShut
  {NoStop}%
\bibitem [{\citenamefont {Lukin}\ \emph {et~al.}(2001)\citenamefont {Lukin},
  \citenamefont {Fleischhauer}, \citenamefont {Cote}, \citenamefont {Duan},
  \citenamefont {Jaksch}, \citenamefont {Cirac},\ and\ \citenamefont
  {Zoller}}]{PhysRevLett.87.037901}%
  \BibitemOpen
  \bibfield  {author} {\bibinfo {author} {\bibfnamefont {M.~D.}\ \bibnamefont
  {Lukin}}, \bibinfo {author} {\bibfnamefont {M.}~\bibnamefont {Fleischhauer}},
  \bibinfo {author} {\bibfnamefont {R.}~\bibnamefont {Cote}}, \bibinfo {author}
  {\bibfnamefont {L.~M.}\ \bibnamefont {Duan}}, \bibinfo {author}
  {\bibfnamefont {D.}~\bibnamefont {Jaksch}}, \bibinfo {author} {\bibfnamefont
  {J.~I.}\ \bibnamefont {Cirac}}, \ and\ \bibinfo {author} {\bibfnamefont
  {P.}~\bibnamefont {Zoller}},\ }\href {\doibase 10.1103/PhysRevLett.87.037901}
  {\bibfield  {journal} {\bibinfo  {journal} {Phys. Rev. Lett.}\ }\textbf
  {\bibinfo {volume} {87}},\ \bibinfo {pages} {037901} (\bibinfo {year}
  {2001})}\BibitemShut {NoStop}%
\bibitem [{\citenamefont {Jaksch}\ \emph {et~al.}(2000)\citenamefont {Jaksch},
  \citenamefont {Cirac}, \citenamefont {Zoller}, \citenamefont {Rolston},
  \citenamefont {C\^ot\'e},\ and\ \citenamefont {Lukin}}]{PhysRevLett.85.2208}%
  \BibitemOpen
  \bibfield  {author} {\bibinfo {author} {\bibfnamefont {D.}~\bibnamefont
  {Jaksch}}, \bibinfo {author} {\bibfnamefont {J.~I.}\ \bibnamefont {Cirac}},
  \bibinfo {author} {\bibfnamefont {P.}~\bibnamefont {Zoller}}, \bibinfo
  {author} {\bibfnamefont {S.~L.}\ \bibnamefont {Rolston}}, \bibinfo {author}
  {\bibfnamefont {R.}~\bibnamefont {C\^ot\'e}}, \ and\ \bibinfo {author}
  {\bibfnamefont {M.~D.}\ \bibnamefont {Lukin}},\ }\href {\doibase
  10.1103/PhysRevLett.85.2208} {\bibfield  {journal} {\bibinfo  {journal}
  {Phys. Rev. Lett.}\ }\textbf {\bibinfo {volume} {85}},\ \bibinfo {pages}
  {2208} (\bibinfo {year} {2000})}\BibitemShut {NoStop}%
\bibitem [{\citenamefont {Wilk}\ \emph {et~al.}(2010)\citenamefont {Wilk},
  \citenamefont {Ga\"{e}tan}, \citenamefont {Evellin}, \citenamefont {Wolters},
  \citenamefont {Miroshnychenko}, \citenamefont {Grangier},\ and\ \citenamefont
  {Browaeys}}]{WilGaeBro10}%
  \BibitemOpen
  \bibfield  {author} {\bibinfo {author} {\bibfnamefont {T.}~\bibnamefont
  {Wilk}}, \bibinfo {author} {\bibfnamefont {A.}~\bibnamefont {Ga\"{e}tan}},
  \bibinfo {author} {\bibfnamefont {C.}~\bibnamefont {Evellin}}, \bibinfo
  {author} {\bibfnamefont {J.}~\bibnamefont {Wolters}}, \bibinfo {author}
  {\bibfnamefont {Y.}~\bibnamefont {Miroshnychenko}}, \bibinfo {author}
  {\bibfnamefont {P.}~\bibnamefont {Grangier}}, \ and\ \bibinfo {author}
  {\bibfnamefont {A.}~\bibnamefont {Browaeys}},\ }\href {\doibase
  10.1103/physrevlett.104.010502} {\bibfield  {journal} {\bibinfo  {journal}
  {Physical Review Letters}\ }\textbf {\bibinfo {volume} {104}},\ \bibinfo
  {pages} {010502+} (\bibinfo {year} {2010})}\BibitemShut {NoStop}%
\bibitem [{\citenamefont {Saffman}(2016)}]{0953-4075-49-20-202001}%
  \BibitemOpen
  \bibfield  {author} {\bibinfo {author} {\bibfnamefont {M.}~\bibnamefont
  {Saffman}},\ }\href {http://stacks.iop.org/0953-4075/49/i=20/a=202001}
  {\bibfield  {journal} {\bibinfo  {journal} {Journal of Physics B: Atomic,
  Molecular and Optical Physics}\ }\textbf {\bibinfo {volume} {49}},\ \bibinfo
  {pages} {202001} (\bibinfo {year} {2016})}\BibitemShut {NoStop}%
\bibitem [{\citenamefont {Zhang}\ \emph {et~al.}(2011)\citenamefont {Zhang},
  \citenamefont {Robicheaux},\ and\ \citenamefont {Saffman}}]{zhang2011Magic}%
  \BibitemOpen
  \bibfield  {author} {\bibinfo {author} {\bibfnamefont {S.}~\bibnamefont
  {Zhang}}, \bibinfo {author} {\bibfnamefont {F.}~\bibnamefont {Robicheaux}}, \
  and\ \bibinfo {author} {\bibfnamefont {M.}~\bibnamefont {Saffman}},\ }\href
  {\doibase 10.1103/PhysRevA.84.043408} {\bibfield  {journal} {\bibinfo
  {journal} {Phys. Rev. A}\ }\textbf {\bibinfo {volume} {84}},\ \bibinfo
  {pages} {043408} (\bibinfo {year} {2011})}\BibitemShut {NoStop}%
\bibitem [{\citenamefont {Topcu}\ and\ \citenamefont
  {Derevianko}(2014)}]{topcu2014Divalent}%
  \BibitemOpen
  \bibfield  {author} {\bibinfo {author} {\bibfnamefont {T.}~\bibnamefont
  {Topcu}}\ and\ \bibinfo {author} {\bibfnamefont {A.}~\bibnamefont
  {Derevianko}},\ }\href {\doibase 10.1103/PhysRevA.89.023411} {\bibfield
  {journal} {\bibinfo  {journal} {Phys. Rev. A}\ }\textbf {\bibinfo {volume}
  {89}},\ \bibinfo {pages} {023411} (\bibinfo {year} {2014})}\BibitemShut
  {NoStop}%
\bibitem [{\citenamefont {Wang}\ \emph {et~al.}(2016)\citenamefont {Wang},
  \citenamefont {Surendran}, \citenamefont {Jose}, \citenamefont {Tran},
  \citenamefont {Herrera}, \citenamefont {Whitlock}, \citenamefont {McLean},
  \citenamefont {Sidorov},\ and\ \citenamefont {Hannaford}}]{WANG20161097}%
  \BibitemOpen
  \bibfield  {author} {\bibinfo {author} {\bibfnamefont {Y.}~\bibnamefont
  {Wang}}, \bibinfo {author} {\bibfnamefont {P.}~\bibnamefont {Surendran}},
  \bibinfo {author} {\bibfnamefont {S.}~\bibnamefont {Jose}}, \bibinfo {author}
  {\bibfnamefont {T.}~\bibnamefont {Tran}}, \bibinfo {author} {\bibfnamefont
  {I.}~\bibnamefont {Herrera}}, \bibinfo {author} {\bibfnamefont
  {S.}~\bibnamefont {Whitlock}}, \bibinfo {author} {\bibfnamefont
  {R.}~\bibnamefont {McLean}}, \bibinfo {author} {\bibfnamefont
  {A.}~\bibnamefont {Sidorov}}, \ and\ \bibinfo {author} {\bibfnamefont
  {P.}~\bibnamefont {Hannaford}},\ }\href {\doibase
  https://doi.org/10.1007/s11434-016-1123-x} {\bibfield  {journal} {\bibinfo
  {journal} {Science Bulletin}\ }\textbf {\bibinfo {volume} {61}},\ \bibinfo
  {pages} {1097 } (\bibinfo {year} {2016})}\BibitemShut {NoStop}%
\bibitem [{\citenamefont {Reichel}\ \emph {et~al.}(1999)\citenamefont
  {Reichel}, \citenamefont {H\"{a}nsel},\ and\ \citenamefont
  {H\"{a}nsch}}]{PhysRevLett.83.3398}%
  \BibitemOpen
  \bibfield  {author} {\bibinfo {author} {\bibfnamefont {J.}~\bibnamefont
  {Reichel}}, \bibinfo {author} {\bibfnamefont {W.}~\bibnamefont {H\"{a}nsel}},
  \ and\ \bibinfo {author} {\bibfnamefont {T.~W.}\ \bibnamefont {H\"{a}nsch}},\
  }\href {\doibase 10.1103/PhysRevLett.83.3398} {\bibfield  {journal} {\bibinfo
   {journal} {Phys. Rev. Lett.}\ }\textbf {\bibinfo {volume} {83}},\ \bibinfo
  {pages} {3398} (\bibinfo {year} {1999})}\BibitemShut {NoStop}%
\bibitem [{\citenamefont {Folman}\ \emph {et~al.}(2000)\citenamefont {Folman},
  \citenamefont {Kr\"uger}, \citenamefont {Cassettari}, \citenamefont {Hessmo},
  \citenamefont {Maier},\ and\ \citenamefont
  {Schmiedmayer}}]{PhysRevLett.84.4749}%
  \BibitemOpen
  \bibfield  {author} {\bibinfo {author} {\bibfnamefont {R.}~\bibnamefont
  {Folman}}, \bibinfo {author} {\bibfnamefont {P.}~\bibnamefont {Kr\"uger}},
  \bibinfo {author} {\bibfnamefont {D.}~\bibnamefont {Cassettari}}, \bibinfo
  {author} {\bibfnamefont {B.}~\bibnamefont {Hessmo}}, \bibinfo {author}
  {\bibfnamefont {T.}~\bibnamefont {Maier}}, \ and\ \bibinfo {author}
  {\bibfnamefont {J.}~\bibnamefont {Schmiedmayer}},\ }\href {\doibase
  10.1103/PhysRevLett.84.4749} {\bibfield  {journal} {\bibinfo  {journal}
  {Phys. Rev. Lett.}\ }\textbf {\bibinfo {volume} {84}},\ \bibinfo {pages}
  {4749} (\bibinfo {year} {2000})}\BibitemShut {NoStop}%
\bibitem [{\citenamefont {Folman}\ \emph {et~al.}(2002)\citenamefont {Folman},
  \citenamefont {Kr\"{u}ger}, \citenamefont {Schmiedmayer}, \citenamefont
  {Denschlag},\ and\ \citenamefont {Henkel}}]{Folman2002Microscopic}%
  \BibitemOpen
  \bibfield  {author} {\bibinfo {author} {\bibfnamefont {R.}~\bibnamefont
  {Folman}}, \bibinfo {author} {\bibfnamefont {P.}~\bibnamefont {Kr\"{u}ger}},
  \bibinfo {author} {\bibfnamefont {J.}~\bibnamefont {Schmiedmayer}}, \bibinfo
  {author} {\bibfnamefont {J.}~\bibnamefont {Denschlag}}, \ and\ \bibinfo
  {author} {\bibfnamefont {C.}~\bibnamefont {Henkel}},\ }in\ \href {\doibase
  10.1016/S1049-250X(02)80011-8} {\emph {\bibinfo {booktitle} {Advances In
  Atomic, Molecular, and Optical Physics}}},\ Vol.~\bibinfo {volume} {48},\
  \bibinfo {editor} {edited by\ \bibinfo {editor} {\bibfnamefont
  {B.}~\bibnamefont {Bederson}}\ and\ \bibinfo {editor} {\bibfnamefont
  {H.}~\bibnamefont {Walther}}}\ (\bibinfo  {publisher} {Academic Press},\
  \bibinfo {year} {2002})\ pp.\ \bibinfo {pages} {263--356}\BibitemShut
  {NoStop}%
\bibitem [{\citenamefont {Reichel}(2002)}]{Reichel2002Microchip}%
  \BibitemOpen
  \bibfield  {author} {\bibinfo {author} {\bibfnamefont {J.}~\bibnamefont
  {Reichel}},\ }\href {\doibase 10.1007/s003400200861} {\bibfield  {journal}
  {\bibinfo  {journal} {Applied Physics B: Lasers and Optics}\ }\textbf
  {\bibinfo {volume} {74}},\ \bibinfo {pages} {469} (\bibinfo {year}
  {2002})}\BibitemShut {NoStop}%
\bibitem [{\citenamefont {Leung}\ \emph {et~al.}(2014)\citenamefont {Leung},
  \citenamefont {Pijn}, \citenamefont {Schlatter}, \citenamefont
  {Torralbo-Campo}, \citenamefont {La~Rooij}, \citenamefont {Mulder},
  \citenamefont {Naber}, \citenamefont {Soudijn}, \citenamefont {Tauschinsky},
  \citenamefont {Abarbanel} \emph {et~al.}}]{leung2014magnetic}%
  \BibitemOpen
  \bibfield  {author} {\bibinfo {author} {\bibfnamefont {V.}~\bibnamefont
  {Leung}}, \bibinfo {author} {\bibfnamefont {D.}~\bibnamefont {Pijn}},
  \bibinfo {author} {\bibfnamefont {H.}~\bibnamefont {Schlatter}}, \bibinfo
  {author} {\bibfnamefont {L.}~\bibnamefont {Torralbo-Campo}}, \bibinfo
  {author} {\bibfnamefont {A.}~\bibnamefont {La~Rooij}}, \bibinfo {author}
  {\bibfnamefont {G.}~\bibnamefont {Mulder}}, \bibinfo {author} {\bibfnamefont
  {J.}~\bibnamefont {Naber}}, \bibinfo {author} {\bibfnamefont
  {M.}~\bibnamefont {Soudijn}}, \bibinfo {author} {\bibfnamefont
  {A.}~\bibnamefont {Tauschinsky}}, \bibinfo {author} {\bibfnamefont
  {C.}~\bibnamefont {Abarbanel}},  \emph {et~al.},\ }\href@noop {} {\bibfield
  {journal} {\bibinfo  {journal} {Review of Scientific Instruments}\ }\textbf
  {\bibinfo {volume} {85}},\ \bibinfo {pages} {053102} (\bibinfo {year}
  {2014})}\BibitemShut {NoStop}%
\bibitem [{\citenamefont {Whitlock}\ \emph {et~al.}(2009)\citenamefont
  {Whitlock}, \citenamefont {Gerritsma}, \citenamefont {Fernholz},\ and\
  \citenamefont {Spreeuw}}]{whitlock2009two}%
  \BibitemOpen
  \bibfield  {author} {\bibinfo {author} {\bibfnamefont {S.}~\bibnamefont
  {Whitlock}}, \bibinfo {author} {\bibfnamefont {R.}~\bibnamefont {Gerritsma}},
  \bibinfo {author} {\bibfnamefont {T.}~\bibnamefont {Fernholz}}, \ and\
  \bibinfo {author} {\bibfnamefont {R.}~\bibnamefont {Spreeuw}},\ }\href@noop
  {} {\bibfield  {journal} {\bibinfo  {journal} {New Journal of Physics}\
  }\textbf {\bibinfo {volume} {11}},\ \bibinfo {pages} {023021} (\bibinfo
  {year} {2009})}\BibitemShut {NoStop}%
\bibitem [{\citenamefont {Wang}\ \emph {et~al.}(2017)\citenamefont {Wang},
  \citenamefont {Tran}, \citenamefont {Surendran}, \citenamefont {Herrera},
  \citenamefont {Balcytis}, \citenamefont {Nissen}, \citenamefont {Albrecht},
  \citenamefont {Sidorov},\ and\ \citenamefont
  {Hannaford}}]{PhysRevA.96.013630}%
  \BibitemOpen
  \bibfield  {author} {\bibinfo {author} {\bibfnamefont {Y.}~\bibnamefont
  {Wang}}, \bibinfo {author} {\bibfnamefont {T.}~\bibnamefont {Tran}}, \bibinfo
  {author} {\bibfnamefont {P.}~\bibnamefont {Surendran}}, \bibinfo {author}
  {\bibfnamefont {I.}~\bibnamefont {Herrera}}, \bibinfo {author} {\bibfnamefont
  {A.}~\bibnamefont {Balcytis}}, \bibinfo {author} {\bibfnamefont
  {D.}~\bibnamefont {Nissen}}, \bibinfo {author} {\bibfnamefont
  {M.}~\bibnamefont {Albrecht}}, \bibinfo {author} {\bibfnamefont
  {A.}~\bibnamefont {Sidorov}}, \ and\ \bibinfo {author} {\bibfnamefont
  {P.}~\bibnamefont {Hannaford}},\ }\href {\doibase 10.1103/PhysRevA.96.013630}
  {\bibfield  {journal} {\bibinfo  {journal} {Phys. Rev. A}\ }\textbf {\bibinfo
  {volume} {96}},\ \bibinfo {pages} {013630} (\bibinfo {year}
  {2017})}\BibitemShut {NoStop}%
\bibitem [{\citenamefont {Choi}\ \emph {et~al.}(2005)\citenamefont {Choi},
  \citenamefont {Guest}, \citenamefont {Povilus}, \citenamefont {Hansis},\ and\
  \citenamefont {Raithel}}]{PhysRevLett.95.243001}%
  \BibitemOpen
  \bibfield  {author} {\bibinfo {author} {\bibfnamefont {J.~H.}\ \bibnamefont
  {Choi}}, \bibinfo {author} {\bibfnamefont {J.~R.}\ \bibnamefont {Guest}},
  \bibinfo {author} {\bibfnamefont {A.~P.}\ \bibnamefont {Povilus}}, \bibinfo
  {author} {\bibfnamefont {E.}~\bibnamefont {Hansis}}, \ and\ \bibinfo {author}
  {\bibfnamefont {G.}~\bibnamefont {Raithel}},\ }\href {\doibase
  10.1103/PhysRevLett.95.243001} {\bibfield  {journal} {\bibinfo  {journal}
  {Phys. Rev. Lett.}\ }\textbf {\bibinfo {volume} {95}},\ \bibinfo {pages}
  {243001} (\bibinfo {year} {2005})}\BibitemShut {NoStop}%
\bibitem [{\citenamefont {Singh}\ \emph {et~al.}(2008)\citenamefont {Singh},
  \citenamefont {Volk}, \citenamefont {Akulshin}, \citenamefont {Sidorov},
  \citenamefont {McLean},\ and\ \citenamefont {Hannaford}}]{singh2008one}%
  \BibitemOpen
  \bibfield  {author} {\bibinfo {author} {\bibfnamefont {M.}~\bibnamefont
  {Singh}}, \bibinfo {author} {\bibfnamefont {M.}~\bibnamefont {Volk}},
  \bibinfo {author} {\bibfnamefont {A.}~\bibnamefont {Akulshin}}, \bibinfo
  {author} {\bibfnamefont {A.}~\bibnamefont {Sidorov}}, \bibinfo {author}
  {\bibfnamefont {R.}~\bibnamefont {McLean}}, \ and\ \bibinfo {author}
  {\bibfnamefont {P.}~\bibnamefont {Hannaford}},\ }\href@noop {} {\bibfield
  {journal} {\bibinfo  {journal} {Journal of Physics B: Atomic, Molecular and
  Optical Physics}\ }\textbf {\bibinfo {volume} {41}},\ \bibinfo {pages}
  {065301} (\bibinfo {year} {2008})}\BibitemShut {NoStop}%
\bibitem [{\citenamefont {Lesanovsky}\ and\ \citenamefont
  {Schmelcher}(2005{\natexlab{a}})}]{PhysRevLett.95.053001}%
  \BibitemOpen
  \bibfield  {author} {\bibinfo {author} {\bibfnamefont {I.}~\bibnamefont
  {Lesanovsky}}\ and\ \bibinfo {author} {\bibfnamefont {P.}~\bibnamefont
  {Schmelcher}},\ }\href {\doibase 10.1103/PhysRevLett.95.053001} {\bibfield
  {journal} {\bibinfo  {journal} {Phys. Rev. Lett.}\ }\textbf {\bibinfo
  {volume} {95}},\ \bibinfo {pages} {053001} (\bibinfo {year}
  {2005}{\natexlab{a}})}\BibitemShut {NoStop}%
\bibitem [{\citenamefont {Lesanovsky}\ and\ \citenamefont
  {Schmelcher}(2005{\natexlab{b}})}]{PhysRevA.72.053410}%
  \BibitemOpen
  \bibfield  {author} {\bibinfo {author} {\bibfnamefont {I.}~\bibnamefont
  {Lesanovsky}}\ and\ \bibinfo {author} {\bibfnamefont {P.}~\bibnamefont
  {Schmelcher}},\ }\href {\doibase 10.1103/PhysRevA.72.053410} {\bibfield
  {journal} {\bibinfo  {journal} {Phys. Rev. A}\ }\textbf {\bibinfo {volume}
  {72}},\ \bibinfo {pages} {053410} (\bibinfo {year}
  {2005}{\natexlab{b}})}\BibitemShut {NoStop}%
\bibitem [{\citenamefont {Schmidt}\ \emph {et~al.}(2007)\citenamefont
  {Schmidt}, \citenamefont {Lesanovsky},\ and\ \citenamefont
  {Schmelcher}}]{0953-4075-40-5-015}%
  \BibitemOpen
  \bibfield  {author} {\bibinfo {author} {\bibfnamefont {U.}~\bibnamefont
  {Schmidt}}, \bibinfo {author} {\bibfnamefont {I.}~\bibnamefont {Lesanovsky}},
  \ and\ \bibinfo {author} {\bibfnamefont {P.}~\bibnamefont {Schmelcher}},\
  }\href {http://stacks.iop.org/0953-4075/40/i=5/a=015} {\bibfield  {journal}
  {\bibinfo  {journal} {Journal of Physics B: Atomic, Molecular and Optical
  Physics}\ }\textbf {\bibinfo {volume} {40}},\ \bibinfo {pages} {1003}
  (\bibinfo {year} {2007})}\BibitemShut {NoStop}%
\bibitem [{\citenamefont {Hezel}\ \emph {et~al.}(2006)\citenamefont {Hezel},
  \citenamefont {Lesanovsky},\ and\ \citenamefont
  {Schmelcher}}]{PhysRevLett.97.223001}%
  \BibitemOpen
  \bibfield  {author} {\bibinfo {author} {\bibfnamefont {B.}~\bibnamefont
  {Hezel}}, \bibinfo {author} {\bibfnamefont {I.}~\bibnamefont {Lesanovsky}}, \
  and\ \bibinfo {author} {\bibfnamefont {P.}~\bibnamefont {Schmelcher}},\
  }\href {\doibase 10.1103/PhysRevLett.97.223001} {\bibfield  {journal}
  {\bibinfo  {journal} {Phys. Rev. Lett.}\ }\textbf {\bibinfo {volume} {97}},\
  \bibinfo {pages} {223001} (\bibinfo {year} {2006})}\BibitemShut {NoStop}%
\bibitem [{\citenamefont {Hezel}\ \emph {et~al.}(2007)\citenamefont {Hezel},
  \citenamefont {Lesanovsky},\ and\ \citenamefont
  {Schmelcher}}]{PhysRevA.76.053417}%
  \BibitemOpen
  \bibfield  {author} {\bibinfo {author} {\bibfnamefont {B.}~\bibnamefont
  {Hezel}}, \bibinfo {author} {\bibfnamefont {I.}~\bibnamefont {Lesanovsky}}, \
  and\ \bibinfo {author} {\bibfnamefont {P.}~\bibnamefont {Schmelcher}},\
  }\href {\doibase 10.1103/PhysRevA.76.053417} {\bibfield  {journal} {\bibinfo
  {journal} {Phys. Rev. A}\ }\textbf {\bibinfo {volume} {76}},\ \bibinfo
  {pages} {053417} (\bibinfo {year} {2007})}\BibitemShut {NoStop}%
\bibitem [{\citenamefont {Mayle}\ \emph
  {et~al.}(2009{\natexlab{a}})\citenamefont {Mayle}, \citenamefont
  {Lesanovsky},\ and\ \citenamefont {Schmelcher}}]{PhysRevA.80.053410}%
  \BibitemOpen
  \bibfield  {author} {\bibinfo {author} {\bibfnamefont {M.}~\bibnamefont
  {Mayle}}, \bibinfo {author} {\bibfnamefont {I.}~\bibnamefont {Lesanovsky}}, \
  and\ \bibinfo {author} {\bibfnamefont {P.}~\bibnamefont {Schmelcher}},\
  }\href {\doibase 10.1103/PhysRevA.80.053410} {\bibfield  {journal} {\bibinfo
  {journal} {Phys. Rev. A}\ }\textbf {\bibinfo {volume} {80}},\ \bibinfo
  {pages} {053410} (\bibinfo {year} {2009}{\natexlab{a}})}\BibitemShut
  {NoStop}%
\bibitem [{\citenamefont {Mayle}\ \emph
  {et~al.}(2009{\natexlab{b}})\citenamefont {Mayle}, \citenamefont
  {Lesanovsky},\ and\ \citenamefont {Schmelcher}}]{PhysRevA.79.041403}%
  \BibitemOpen
  \bibfield  {author} {\bibinfo {author} {\bibfnamefont {M.}~\bibnamefont
  {Mayle}}, \bibinfo {author} {\bibfnamefont {I.}~\bibnamefont {Lesanovsky}}, \
  and\ \bibinfo {author} {\bibfnamefont {P.}~\bibnamefont {Schmelcher}},\
  }\href {\doibase 10.1103/PhysRevA.79.041403} {\bibfield  {journal} {\bibinfo
  {journal} {Phys. Rev. A}\ }\textbf {\bibinfo {volume} {79}},\ \bibinfo
  {pages} {041403R} (\bibinfo {year} {2009}{\natexlab{b}})}\BibitemShut
  {NoStop}%
\bibitem [{\citenamefont {Gerritsma}\ and\ \citenamefont
  {Spreeuw}(2006)}]{PhysRevA.74.043405}%
  \BibitemOpen
  \bibfield  {author} {\bibinfo {author} {\bibfnamefont {R.}~\bibnamefont
  {Gerritsma}}\ and\ \bibinfo {author} {\bibfnamefont {R.~J.~C.}\ \bibnamefont
  {Spreeuw}},\ }\href {\doibase 10.1103/PhysRevA.74.043405} {\bibfield
  {journal} {\bibinfo  {journal} {Phys. Rev. A}\ }\textbf {\bibinfo {volume}
  {74}},\ \bibinfo {pages} {043405} (\bibinfo {year} {2006})}\BibitemShut
  {NoStop}%
\bibitem [{\citenamefont {Davis}(2002)}]{Davis2002}%
  \BibitemOpen
  \bibfield  {author} {\bibinfo {author} {\bibfnamefont {T.}~\bibnamefont
  {Davis}},\ }\href {\doibase 10.1140/e10053-002-0003-x} {\bibfield  {journal}
  {\bibinfo  {journal} {The European Physical Journal D - Atomic, Molecular,
  Optical and Plasma Physics}\ }\textbf {\bibinfo {volume} {18}},\ \bibinfo
  {pages} {27} (\bibinfo {year} {2002})}\BibitemShut {NoStop}%
\bibitem [{\citenamefont {Naber}(2016)}]{naber2016magnetic}%
  \BibitemOpen
  \bibfield  {author} {\bibinfo {author} {\bibfnamefont {J.~B.}\ \bibnamefont
  {Naber}},\ }\emph {\bibinfo {title} {Magnetic atom lattices for quantum
  information}},\ \href {http://hdl.handle.net/11245/1.543967} {Ph.D. thesis},\
  \bibinfo  {school} {University of Amsterdam} (\bibinfo {year} {2016}),\
  \bibinfo {note} {see chapter 2, section 2.3.2}\BibitemShut {NoStop}%
\bibitem [{Note1()}]{Note1}%
  \BibitemOpen
  \bibinfo {note} {We have neglected a number of terms in Eq. \protect \textup
  {\hbox {\mathsurround \z@ \protect \normalfont (\ignorespaces \ref
  {eq:HRr}\unskip \@@italiccorr )}}, part of the perturbation Hamiltonian
  \begin {align*} H&_\protect \mathrm {small}= \protect \bm {S}\cdot \protect
  \mathaccentV {tilde}07E{\protect \bm {B}}_q(\protect \bm {R},\protect \bm
  {r}) +\protect \frac {\alpha ^2}{2r}\protect \frac {dV_l(r)}{dr}(\protect \bm
  {A}(\protect \bm {R})\times \protect \bm {r}) \cdot \protect \bm {S}\\
  +&\protect \frac {i}{2}\protect \mathcal {G}{\setbox \z@ \hbox
  {\frozen@everymath \@emptytoks \mathsurround \z@ $\nulldelimiterspace \z@
  \left (\vcenter to\@ne \big@size {}\right .$}\box \z@ }X\left [H_\protect
  \mathrm {ff},yz\right ] +Y\left [H_\protect \mathrm {ff},xz\right ]{\setbox
  \z@ \hbox {\frozen@everymath \@emptytoks \mathsurround \z@
  $\nulldelimiterspace \z@ \left )\vcenter to\@ne \big@size {}\right .$}\box
  \z@ }\\ +&\protect \frac {i}{2}\partial ^R_j\protect \bm {A}_{q,k} (\protect
  \bm {R})\left [H_\protect \mathrm {ff},r_jr_k\right ] +\protect \bm
  {A}^{(2)}_q(\protect \bm {R},\protect \bm {r})\cdot \protect \bm {p},\label
  {eq:negl} \end {align*} which have all been estimated to give only minor
  contributions. With the exception of $\protect \bm {S}\cdot \protect
  \mathaccentV {tilde}07E{\protect \bm {B}}_q(\protect \bm {R},\protect \bm
  {r})$ term, all have higher order than 3 in the relative
  coordinates.}\BibitemShut {Stop}%
\bibitem [{\citenamefont {Wineland}\ \emph {et~al.}(1998)\citenamefont
  {Wineland}, \citenamefont {Monroe}, \citenamefont {Itano}, \citenamefont
  {Leibfried}, \citenamefont {King},\ and\ \citenamefont
  {Meekhof}}]{Wineland98}%
  \BibitemOpen
  \bibfield  {author} {\bibinfo {author} {\bibfnamefont {D.~J.}\ \bibnamefont
  {Wineland}}, \bibinfo {author} {\bibfnamefont {C.}~\bibnamefont {Monroe}},
  \bibinfo {author} {\bibfnamefont {W.~M.}\ \bibnamefont {Itano}}, \bibinfo
  {author} {\bibfnamefont {D.}~\bibnamefont {Leibfried}}, \bibinfo {author}
  {\bibfnamefont {B.~E.}\ \bibnamefont {King}}, \ and\ \bibinfo {author}
  {\bibfnamefont {D.~M.}\ \bibnamefont {Meekhof}},\ }\href {\doibase
  10.6028/jres.103.019} {\bibfield  {journal} {\bibinfo  {journal} {Journal of
  Research of the National Institute of Standards and Technology}\ }\textbf
  {\bibinfo {volume} {103}},\ \bibinfo {pages} {259} (\bibinfo {year}
  {1998})}\BibitemShut {NoStop}%
\bibitem [{\citenamefont {Beterov}\ \emph {et~al.}(2009)\citenamefont
  {Beterov}, \citenamefont {Ryabtsev}, \citenamefont {Tretyakov},\ and\
  \citenamefont {Entin}}]{PhysRevA.79.052504}%
  \BibitemOpen
  \bibfield  {author} {\bibinfo {author} {\bibfnamefont {I.~I.}\ \bibnamefont
  {Beterov}}, \bibinfo {author} {\bibfnamefont {I.~I.}\ \bibnamefont
  {Ryabtsev}}, \bibinfo {author} {\bibfnamefont {D.~B.}\ \bibnamefont
  {Tretyakov}}, \ and\ \bibinfo {author} {\bibfnamefont {V.~M.}\ \bibnamefont
  {Entin}},\ }\href {\doibase 10.1103/PhysRevA.79.052504} {\bibfield  {journal}
  {\bibinfo  {journal} {Phys. Rev. A}\ }\textbf {\bibinfo {volume} {79}},\
  \bibinfo {pages} {052504} (\bibinfo {year} {2009})}\BibitemShut {NoStop}%
\bibitem [{\citenamefont {S\o{}rensen}\ and\ \citenamefont
  {M\o{}lmer}(1999)}]{PhysRevLett.82.1971}%
  \BibitemOpen
  \bibfield  {author} {\bibinfo {author} {\bibfnamefont {A.}~\bibnamefont
  {S\o{}rensen}}\ and\ \bibinfo {author} {\bibfnamefont {K.}~\bibnamefont
  {M\o{}lmer}},\ }\href {\doibase 10.1103/PhysRevLett.82.1971} {\bibfield
  {journal} {\bibinfo  {journal} {Phys. Rev. Lett.}\ }\textbf {\bibinfo
  {volume} {82}},\ \bibinfo {pages} {1971} (\bibinfo {year}
  {1999})}\BibitemShut {NoStop}%
\bibitem [{\citenamefont {S\o{}rensen}\ and\ \citenamefont
  {M\o{}lmer}(2000)}]{PhysRevA.62.022311}%
  \BibitemOpen
  \bibfield  {author} {\bibinfo {author} {\bibfnamefont {A.}~\bibnamefont
  {S\o{}rensen}}\ and\ \bibinfo {author} {\bibfnamefont {K.}~\bibnamefont
  {M\o{}lmer}},\ }\href {\doibase 10.1103/PhysRevA.62.022311} {\bibfield
  {journal} {\bibinfo  {journal} {Phys. Rev. A}\ }\textbf {\bibinfo {volume}
  {62}},\ \bibinfo {pages} {022311} (\bibinfo {year} {2000})}\BibitemShut
  {NoStop}%
\bibitem [{\citenamefont {Wang}\ \emph {et~al.}(2001)\citenamefont {Wang},
  \citenamefont {S\o{}rensen},\ and\ \citenamefont
  {M\o{}lmer}}]{PhysRevLett.86.3907}%
  \BibitemOpen
  \bibfield  {author} {\bibinfo {author} {\bibfnamefont {X.}~\bibnamefont
  {Wang}}, \bibinfo {author} {\bibfnamefont {A.}~\bibnamefont {S\o{}rensen}}, \
  and\ \bibinfo {author} {\bibfnamefont {K.}~\bibnamefont {M\o{}lmer}},\ }\href
  {\doibase 10.1103/PhysRevLett.86.3907} {\bibfield  {journal} {\bibinfo
  {journal} {Phys. Rev. Lett.}\ }\textbf {\bibinfo {volume} {86}},\ \bibinfo
  {pages} {3907} (\bibinfo {year} {2001})}\BibitemShut {NoStop}%
\end{thebibliography}
\end{document}